\documentclass[a4paper,fleqn,usenatbib]{mnras}
\usepackage{newtxtext,newtxmath}

\usepackage[T1]{fontenc}
\usepackage{ae,aecompl}

\usepackage{graphicx}	
\usepackage{amsmath}	
\usepackage{amssymb}	
\usepackage{mathtools}
\usepackage{hyperref}

\newcommand{\note}[1]{\textcolor{black}{#1}}

\title[The Maximum Mass Solar Nebula]{The Maximum Mass Solar Nebula and the early formation of planets}

\author[Nixon et al.]{
C. J. Nixon$^{1}$\thanks{E-mail: \href{mailto:cjn@leicester.ac.uk}{cjn@leicester.ac.uk}},
A. R. King$^{1,2,3}$ and
J. E. Pringle$^{1,4}$\\
$^{1}$Theoretical Astrophysics Group, Department of Physics \& Astronomy, University of Leicester, Leicester, LE1 7RH\\
$^{2}$Anton Pannekoek Institute, University of Amsterdam, Science Park 904, 1098 XH Amsterdam, Netherlands\\ 
$^{3}$Leiden Observatory, Leiden University, Niels Bohrweg 2, NL-2333 CA Leiden, Netherlands\\
$^{4}$Institute of Astronomy, Madingley Road, Cambridge, CB3 0HA
}

\date{Draft version, \today.}
\pubyear{2017}
\begin{document}
\label{firstpage}
\pagerange{\pageref{firstpage}--\pageref{lastpage}}
\maketitle

\begin{abstract}
Current planet formation theories provide successful frameworks with which to interpret the array of new observational data in this field. However, each of the two main theories (core accretion, gravitational instability) is unable to explain some key aspects. In many planet formation calculations, it is usual to treat the initial properties of the planet forming disc (mass, radius, etc.) as free parameters. In this paper, we stress the importance of setting the formation of planet forming discs within the context of the formation of the central stars. By exploring the early stages of disc formation, we introduce the concept of the Maximum Mass Solar Nebula (MMSN), as opposed to the oft-used Minimum Mass Solar Nebula (here mmsn). It is evident that almost all protoplanetary discs start their evolution in a strongly self-gravitating state. In agreement with almost all previous work in this area, we conclude that on the scales relevant to planet formation these discs are not gravitationally unstable to gas fragmentation, but instead form strong, transient spiral arms. These spiral arms can act as efficient dust traps allowing the accumulation and subsequent fragmentation of the dust (but not the gas). This phase is likely to populate the disc with relatively large planetesimals on short timescales while the disc is still veiled by a dusty-gas envelope. Crucially, the early formation of large planetesimals overcomes the main barriers remaining within the core accretion model. A prediction of this picture is that essentially all observable protoplanetary discs are already planet hosting.
\end{abstract}


\section{Introduction}

Planets form in discs around their host stars. These discs are approximately Keplerian, and moderately thin with aspect ratios $H/R \sim 0.1$. Through the lifetime of the disc, of a few million years, much of the disc mass drains on to the central protostar, while some condenses into planets (some of which can be ejected or accreted by the central star) and some is lost to outflows (e.g. protostellar jets, photoevaporative winds and magnetic winds).

Broadly speaking, there are two prevalent planet formation scenarios: (1) core accretion \citep[e.g.][]{Mizuno:1980aa,Lissauer:1993aa,Pollack:1996aa}, where planetesimals grow from dust within the disc and coagulate to form rocky planets, with subsequent accretion of gas to create gas giants, and (2) gravitational instability \citep[e.g.][]{Adams:1989aa,Boss:1997aa,Durisen:2007aa,Nayakshin:2017aa}, where the protoplanetary gas disc fragments into gaseous clumps. These subsequently accrete solids from the disc.

Both of these scenarios have had success in predicting and explaining observed properties of exoplanet systems, but neither provides a complete picture of planet formation. Each scenario has major shortcomings which are still yet to be resolved. For example within core accretion, among $\sim 10$\,cm sized \note{particles} the higher relative velocity \citep{Weidenschilling:1993aa} of collisions results in fragmentation rather than growth of planetesimals \citep{Blum:2008aa}. Further, at sizes $\sim 1$\,m, \note{particles} can spiral into the central star through aerodynamical drag on short timescales \citep[$\sim 100$\,yrs;][]{Weidenschilling:1977aa} faster than the planet formation process is expected to proceed. At larger sizes the protoplanets can also migrate rapidly through the disc on to the central star -- the Type I migration problem. Within gravitational instability models, only the outer regions of large and massive protoplanetary discs are thought to fragment, leading to the formation of massive gas giants and/or brown dwarfs \citep[e.g.][]{Kratter:2006aa}. We note that fragmentation into multiple-star systems on these scales has been directly observed \citep{Tobin:2016aa}. Any planets formed through this route are already close to the brown dwarf limit, and as they migrate they grow in mass (reaching $\gtrsim$ brown dwarf masses) until the disc is unable to force significant further migration \citep{Stamatellos:2015aa}. Also there appears to be a discrepancy in composition of planets formed by gravitational instability, with the planets theoretically expected to have near-solar metallicity \citep[e.g.][]{Helled:2010aa}, but inferences from observations suggest factors $10-100$ larger \cite[e.g.][]{Thorngren:2016aa}. \note{Recently \cite{Ilee:2017aa} have shown that the composition of planets formed by gravitational instability may also depend upon their dynamical history.} All of these issues in both models are complex and, with different initial conditions or other variants, can be ameliorated to a degree. But it is clear that no complete picture exists in either model.

There is also recent observational evidence that discs form early and are massive enough for self-gravity to play an important role. There are observations of gas rotation in young (Class 0/I) protostars \citep{Tobin:2012aa,Tobin:2015aa,Yen:2015aa}. ALMA has recently spatially resolved several protoplanetary discs \citep[e.g.][]{ALMA-Partnership:2015aa,Perez:2016aa,Tobin:2016aa}. In particular \cite{Perez:2016aa} observe structures consistent with gravitational spiral arms \note{(see also \citealt{Meru:2017aa,Hall:2018aa})}. In this system the disc mass is $\sim 0.1M_\odot$ and the spiral arms are at $100-300$\,AU.

Thus a general picture is emerging from observations that the planet formation process is vigorous, with almost all stars likely to host planets \citep[e.g.][]{Fressin:2013aa}. \cite{Winn:2015aa} note that a sun-like star has a $10$ per cent chance of hosting a giant planet with orbital period less than a few years, and a $\sim 50$ per cent chance of hosting a compact smaller planetary system. This picture is reinforced by the fact that the solar system is dynamically full \citep[see the discussion in Section 4.2 of][]{Laskar:1996aa} and therefore may have produced many more planets than are remaining today --  the ones that remain are marginally stable on the solar system's lifetime, and any that were not would have been ejected. This thesis, that planet formation is plentiful and wasteful, is further supported by \cite{Zinzi:2017aa} who analyse the multiplicity and angular momentum deficit of exoplanet systems, finding an anti-correlation which leads them to conclude that exoplanet systems can become unstable and lose some of their planets.

\note{Additionally}, significant numbers of planets may have been accreted on to the central star by migrating through the disc\note{, further reducing the efficiency of the planet formation process}. \cite{Armitage:2002aa} \note{explore this possibility, finding} that what is left is created by the critical point where migration becomes inefficient (presumably as the disc mass decays).\footnote{It is plausible that the accretion of several planets can provide a solution to the luminosity problem that protostars are typically less bright than expected based on the time required to accrete their mass \citep{Kenyon:1990aa}. However, we note that this could also be explained by FU Orionis type outburst cycles driven by e.g. the gravito-magnetic cycle of \cite{Martin:2011aa,Martin:2012aa}.}

The inference we make from this is that to understand planet formation in general we must place it in the context of the formation of the protoplanetary disc, as this sets the initial conditions for planet formation. This in turn means placing disc formation in the context of the formation of the central protostar. We also stress that in terms of mass, accounting for the formation of Jupiter is 90 per cent of the picture within the solar system. Forming this first giant planet is difficult as migration timescales and radial drift of dust can be rapid, but once Jupiter is formed and opens a gap in the disc gas, Type II migration is considerably slower. Thus forming gas giants like Jupiter from a protoplanetary disc may be the important step.

In this work we review the current ideas on the initial formation of planet forming discs, and infer typical properties of these discs. We find that, in general, predictions for the mass of these discs at early times (approximately during the early protostar phase) are that they contain significant fractions of the host star mass ($M_{\rm disc}/M_\star \sim 0.1-$a few). This has two implications. First very young discs, where there is still abundant material for planet formation, are strongly self-gravitating\footnote{By self-gravitating, we mean that the local gas self-gravity from the disc is dynamically important. Whether such a disc fragments or forms spiral arms depends also on the cooling rate. We discuss this dynamics in more detail later.}. Second, it follows that the oft-used concept of a minimum mass solar nebula (mmsn) is an unlikely starting condition for planet-formation. We suggest that a more useful concept is that of the Maximum Mass Solar Nebula, which we will refer to here as MMSN.

\section{Initial conditions for planet formation}

The initial conditions for planet formation calculations are often described in terms of a minimum mass solar nebula (mmsn). These calculations have a long history, dating back to e.g. \cite{Kuiper:1956aa,Kusaka:1970aa,Edgeworth:1949aa,Safronov:1967aa,Alfven:1970aa,Lecar:1973aa}. \cite{Weidenschilling:1977ab} coined the term mmsn, and adding the solar complement of light elements to the planet masses obtained a value of  approximately $\approx 10 M_{\rm J} \approx 0.01 M_\odot$. Since these early works a variety of processes at work in protoplanetary disc evolution have been discovered, and many of these result in the removal of significant mass from the disc, for example: 1. accretion on to the star, 2. photoevaporation/winds, 3. ejection of formed planets, and 4. subsequent accretion of formed planets. As such it is a surprise that the concept of a mmsn is still used in the consideration of initial conditions for planet forming discs \cite[e.g.][]{Simon:2015aa,Lines:2015aa,Bitsch:2015aa,Hopkins:2016aa,Coleman:2017aa,Mutter:2017aa}\footnote{\note{\cite{Chiang:2013aa} construct a minimum mass extrasolar nebula of solar composition solids and gas by inferring the disc surface density from the distribution of planets observed by {\it Kepler}. This methodology contains the same issues we raise here.}}. It is worth stressing that adoption of the mmsn as the initial starting point for a planet forming disc implies  $ \approx 100$ per cent efficiency for the planet formation process. In view of this, many investigations use several, up to around $10$, times this value (i.e. up to $\sim 100 M_{\rm J} \sim 0.1M_{\odot}$). In these cases the disc mass is $\lesssim 0.01-0.1 M_\odot$, which for a solar type star would result in a non-self-gravitating disc, with $M_{\rm disc}/M_\star \lesssim H/R$.

Thus, when the concept of a  mmsn is invoked, the efficiency of planet formation is assumed to be $>10-100$ per cent. This seems optimistic. If we instead assumed say $1-10$ per cent efficiency, we would need a disc with a mass $M_{\rm disc}/M_\star \approx 0.1 - 1$. Since these values are comparable to or greater than the dimensionless disc thickness $H/R$ this implies we would expect the disc self-gravity to be important \citep{Pringle:1981aa}. This is particularly likely if we account for the mass lost through the processes described above. All of these points render the idea of taking a mmsn (or a small multiple thereof) as a starting model, as potentially futile.

In the next two sub-sections we discuss the formation of planet forming discs and their early-time properties within the context of star formation. Taking the lead from the star formation literature, we split this discussion into two parts, first a single-star monolithic collapse scenario, and second a more dynamic scenario including binary/triple etc star formation and possible subsequent ejection of planet forming stellar systems.

\subsection{Monolithic collapse star formation - simple disc formation and evolution}

Early ideas about star formation centred on the concept of an isolated, rotating, collapsing protostellar core \citep[e.g.][]{Shu:1987aa}. Here we describe some of the previous findings which considered the initial disc formation and evolution, starting from an isolated ``core'', which is assumed rotating. The properties of the core ensure a surface accretion rate onto a disc, with the later gas to arrive having more angular momentum, and then the disc evolution is considered with some simple approximations.  In all these computations the mass of the star starts in the core, and is then processed through a disc, onto the star. As we proceed, we bear in mind the question: when does star formation cease and planet formation begin?
\begin{enumerate}
\item \cite{Lin:1990aa} consider an accreting core of mass 1.0 $M_\odot$, with a free fall timescale of $t_{ff} = 2 \times 10^5$\,yr which forms a star and surrounding disc. The disc evolution is followed using a \cite{Shakura:1973aa} $\alpha$ and where necessary using an $\alpha$ to mimic the effects of self-gravity \citep[][\note{see also \citealt{Rice:2010aa} who perform similar calculations and find similar results}]{Lin:1987aa}. They report various models with disc sizes of 700 AU and 1700 AU for which self-gravity is relevant,  typically finding that the mass starts mostly in the disc and then drains onto the star. For the model in which the accretion occurs out to $a_{\rm max} = 1700$AU, at the end of the computation at a time $4 \times 10^6$\,yr, they find $M_{\rm disc}/M_\star \approx 1.5$. Thus the disc is massive and large with a maximum disc mass comparable to the stellar mass.

\item \cite{Nakamoto:1995aa} do not consider self-gravity of the discs and use an $\alpha$-viscosity. They consider an idealised rotating core which gives rise to an accretion rate onto the disc. In their Fig.~5 they give the results for various values of $\alpha$ and the core's angular momentum $J$. Typical final values are in the range $M_{\rm disc}/M_\star \approx 0.05 - 2$.

\item \cite{Walch:2009aa} consider the collapse of rotating cores using SPH in 3D. They are interested in looking for fragmentation of the disc as it forms in order to look at formation of planets. Their disc's sizes are in the range $R_{\rm disc} \approx 100 - 1,000$\,AU. The larger ones can lead to fragments, but typically the effect of self-gravity is to induce the formation of spiral structure and the rapid transfer of angular momentum. In Fig. 4 they give maximum disc masses of $M_{\rm disc}/M_\star \approx 2 - 5$. 

\item \cite{Jin:2010aa} consider the collapse of a $1 - 2 M_\odot$ core  to discs with radii $R_{\rm disc} \sim 100$AU. They assume $\alpha = 0.02$ when the disc is self-gravitating. In their Fig. 4 they show the `standard case' which has a small disc (most mass within 20 AU) and $\alpha = 0.0015$. For this case they find a maximum value of $M_{\rm disc}/M_\star \approx 0.08$.

\item \cite{Tomida:2017aa} aimed to match Elias 2-27 with a long term MHD simulation using a 3D nested-grid code. Their initial conditions contained a super-critical Bonnor-Ebert sphere, with T = 10 K, $R = 6.1 \times 10^3$ AU and $M = 1.25 M_\odot$. They assumed solid body rotation and an initial $B_z = 36 \mu$G. As accretion proceeds, the disc steadily grows, and goes through a succession of strong self-gravitational events, in which grand design spiral arms are seen. The disc radius steadily grows, but with strong fluctuations, reaching a radius of around 150 -- 200 AU. The accretion rate is around $10^{-5} M_\odot$/yr. The simulation is carried out for 50,000\,yr, by which time most of the envelope has fallen in. Half of the accreted material is lost to an outflow. Throughout the  simulation the disc mass and stellar mass have a ratio of $M_{\rm disc}/M_\star \approx 0.5$.
\end{enumerate}

\subsubsection{Summary}

These are all single core monolithic collapse calculations in the style of the \cite{Shu:1987aa} view of star formation. Depending on the assumed angular momentum of the collapsing material one gets discs of different sizes. The larger discs take longer to let the material accrete onto the central stars and so will tend to be more massive. As we have seen, all of these calculations result in discs that are massive enough that self-gravity is dynamically important with $M_{\rm disc}/M_\star \approx 0.1-{\rm a~few}$, and disc radii of $100-1000$\,AU.

The question we are trying to answer is: when is the most appropriate moment to start considerations of planet formation, and what are the disc conditions then? In this picture of single core collapse, the simple answer appears to be to assume that the initial conditions at which consideration of planet formation starts to take place should be when the disc reaches its  maximum mass. In this case, as we have suggested, it makes sense to replace the concept of the minimum mass solar nebula (mmsn) with one of the Maximum Mass Solar Nebula (MMSN). It is then clear that at this point the disc is self-gravitating.

From the point of view of star formation the monolithic collapse scenario always leads to the formation of single stars. This is not a good result for solar mass stars of which only $\approx 50 \pm 10$ per cent are single \citep{Raghavan:2010aa}. Thus this scenario cannot be the whole picture. However, from the point of view of planet formation in the context of currently observed planets this picture might be more relevant as the focus has been on finding planets around solar mass, single stars \citep{Winn:2015aa}.

\subsection{Planet forming conditions in the context of the current star formation picture}

Almost all planets are currently observed around single stars. It is not yet clear whether this is an observational bias or that it is harder to form them in binaries. We know that most solar-type stars are not singles, but are, rather, members of binary or multiple systems \citep{Raghavan:2010aa}. Therefore we have to ask how most single stars form, and to build a planet formation picture in this view.

In an effort to explain the large fraction of multiple star systems, \cite{Pringle:1989aa,Pringle:1991aa,Clarke:1991aa} make the case that, in contrast to the \cite{Shu:1987aa} picture, to understand the formation of single stars it is necessary to understand the formation of binary and multiple stars. This is because in order to account for the number of binary and multiple systems it is necessary that essentially {\it all} stars have to form with friends. If all stars form in groups, then some of these will be ejected as single stars. And given that single stars are a minority, it follows that most stars must form in groups. These ideas are crystallised in \cite{Bate:2003aa,Bate:2009aa,Bate:2012aa}. The importance of binary formation on the monolithic collapse picture given above is evident from the fact that the median binary separation for solar mass stars is around $20-40$\,AU (see Fig.~7 of \citealt{Duquennoy:1991aa} and Fig.~13 of \citealt{Raghavan:2010aa}). This implies that the dynamical interactions that take place during the formation process are likely to play a role in determining disc sizes, disc evolution and therefore disc masses.

{\bf Initial conditions:} The simulation described by \cite{Bate:2012aa} produces a good fit to the stellar mass function in the range $0.02 \le M/M_\odot \le 3$, and also produces good fits to the binary, and multiple, star fractions as well as to the properties of multiple stars such as periods (separations) etc and so we focus on this here. These simulations start with an isolated, self-gravitating, uniform sphere of gas, of mass 500 $M_\odot$ and radius 0.404 pc. Thus the mean density is $\bar{\rho} = 1.3 \times 10^{-19}$ g/cm$^3$, equivalent to a hydrogen molecule density of $n(H_{II}) = 4 \times 10^4$ cm$^{-3}$. The initial free-fall time of the cloud is $t_{ff} = 1.9 \times 10^5$yr, and the initial Jeans mass in the cloud is $\approx 1 M_\odot$. The cloud is a bit more compact than might be deemed typical for nearby (reasonably low mass, low density) star forming regions, and is more representative of the denser regions in more distant, more massive molecular clouds in which the majority of star formation is expected to take place. For our purposes here, the main worry is that the cloud is isolated and self-gravitating, and so most of the gas must end up as stars, leading to a much higher star formation efficiency than is globally reasonable. This is in contrast to the current view that molecular clouds are transient events produced by convergent flows in/near spiral arms \citep{Dobbs:2006aa,Dobbs:2011aa,Dobbs:2012aa}. However, at the very least, this simulation can give us an idea of the kind of physical processes we need to be considering when we want to think about the initial stages of planet-forming discs.

The simulations are initiated with a temperature of 10K, but with a substantial turbulent velocity spectrum with mean Mach number of around 14. The collapse is then followed, using sink particles with radii of $5.0$\,AU ($0.5$\,AU was also simulated, but not ran as long).\footnote{The simulations of \cite{Krumholz:2011aa} produce substantial agreement but have sink particle sizes of $>100$\,AU, and are therefore unable to resolve the regions we require for discussion here.}

{\bf Initial disc masses:} The mass function that emerges from these calculations (and which agrees well with observational data; \citealt{Bate:2012aa}) seems to be caused predominantly by the time between formation of a star (here a sink particle) and termination of accretion onto it. The accretion rates onto growing objects are typically around $1 - 2 \times 10^{-5} M_\odot$/yr. The termination of accretion seems to be caused (at least for single stars) by a dynamical interaction which ejects the star from the denser parts of the cloud. Looking at the detailed structure of discs in these conditions is problematic. The mass of an SPH particle is $m_P \approx 10^{-5} M_\odot \approx 0.01 M_J$. Thus a mmsn disc has $\sim 10^3$ particles, and a reasonable self-gravitating disc has around $10^4$ particles. This is neither enough to resolve disc structure nor to accurately resolve disc evolution. So, most likely, the disc masses when accretion stops may not be accurate. \cite{Bate:2009aa,Bate:2012aa} provides an estimate of the disc size when accretion stops. This is given in Figs. 14 of \cite{Bate:2009aa} and 15 of \cite{Bate:2012aa}. What is plotted here is one half of the ``closest encounter distance''. The time at which this encounter occurs is identified by when the last large acceleration occurred (defined as greater than 2000 km/s/Myr).

In Fig. 15 of \cite{Bate:2012aa}, we find that for all the stars with $M < 0.3 M_\odot$, around one half of them have a ``last encounter'' half-distance of $> 10$\,AU, and about 5 per cent have $> 100$\,AU. The stars in the range $0.3 < M/M_\odot < 1.0$ are quite different. This may be because the binary fraction is largest in this range, so that many of the closest encounters in this range are because the stars are binaries.  This means that the most massive stars (the ones that stay longest in the high density gas) are still accreting at the end of the simulation. In addition, the dynamics of binaries and multiple systems can take a while to sort out, not least because some of those stars apparently ejected are not totally unbound and so can return to the fray at a later date. And \cite{Bate:2009aa,Bate:2012aa} warns that these numbers need to be treated with caution -- for example, a star can suffer a close dynamical encounter and then accrete some more material from the cloud. It is also true that because (for numerical reasons) the initial cloud is chosen to be denser than optimal, the last closest encounters might well be smaller than is usual. Since disc evolution timescales scale strongly with radius, it might well be that these simulations would tend to underestimate initial disc masses. 

With these caveats in mind, \cite{Bate:2018aa} provides a more detailed analysis of the disc properties in his simulation. He underscores the point that in this scenario the chaotic nature of the star formation process gives rise to an enormous diversity in the properties of the very young protostellar discs that are present at the end of the simulation. He finds disc radii in the range of order ten to a few hundred AU, and finds typical maximum disc to star mass ratios in the range $M_{\rm disc}/M_\star \sim 0.1 - 1$.

\subsubsection{Summary}

Thus we find that all of the simulations of early disc formation in the context of star formation imply that the idea of  the simplest planet-forming initial conditions might have some meaning. We note that here we need to consider mainly discs around single stars since this is where most planets are found, presumably for reasons of selection.  In this case, in the more realistic scenario of chaotic star formation,  it seems reasonable to adopt the concept of an ``initial disc mass'' (and also ``initial disc radius''), which is the mass when accretion stops, perhaps caused by a strong dynamical interaction. However, as we have seen, to get a good handle on the properties of such initial discs requires much more detailed, higher-resolution simulations

In all cases, however,  the discs are found to be a large fraction of the central star's mass, indicating the disc self-gravity plays an important role. At the time this is happening, the disc is shrouded by a dusty envelope (Class 0/I protostar). This implies that observing this phase may be difficult, if not  impossible. All of the above points towards taking as initial conditions a Maximum Mass Solar Nebula (MMSN), which from many starting points appears to be of order the central star's mass. At this stage the disc is self-gravitating.  Thus while the general picture in the binary/multiple star formation scenario is more complex than the single, isolated core case, the initial conditions for planet formation are similar: the discs are massive and self-gravitating.

\section{Subsequent evolution and planet formation}

We have discussed above the current understanding of the formation of protoplanetary discs. Whether this occurs in isolation or in molecular clouds with dynamical interactions etc, the assembly of the star is contemporaneous with the assembly of the disc. This implies that we need be thinking about star and planet formation as one continuous process. This point is also made by \cite{Kuffmeier:2017aa}. The current paradigm is to assume a central star exists with a disc around it, but with the properties of the star and disc disconnected. This has prompted the misleading treatment of  concepts such as  ``initial disc mass'' in terms of the mmsn, and "initial disc size" as if they were free parameters.

We suggest that a reasonable initial condition is to start when the accretion rate onto the disc drops (i.e. the MMSN for the monolithic collapse scenario) or when the star is expelled from the accretion environment (i.e. MMSN or Initial Disc Mass, for the more realistic concept of the dynamical star-forming environment). As most of the mass in many planetary systems is located in the higher mass gas giants, we should consider their formation at this stage.

The main property of these initial MMSN discs is that they are self-gravitating. Nearly all investigations find that such discs ($\sim 10-100$ per cent of the stellar mass) do not fragment. Due to the long cooling times in protostellar discs, the many, short-lived transient spiral arms that form from the gravitational instability drive efficient angular momentum transfer rather than fragmentation into clumps. Simulations which do find fragmentation \citep[e.g.][]{Hall:2017aa} start with a ready-formed strongly gravitating disc. Building the disc up over time changes the picture. Even in this limiting case of starting with a self-gravitating disc, the general picture is that the planets we observe and are trying to explain simply do not form \citep[e.g.][]{Rafikov:2005aa,Kratter:2006aa,Forgan:2017aa}. Instead wide separation, low-mass stars are the typical end product \citep{Kratter:2016aa}.

We therefore need to consider  self-gravitating protoplanetary discs whose angular momentum transport is principally driven through self-gravity. The nature of the self-gravitational behaviour is to gather the solids into clumps on a near-dynamical timescale. \cite{Rice:2004aa,Rice:2006aa} show that the spiral arm formation can provide efficient dust traps, creating large dust densities required to produce planetesimals $\gtrsim$ metres, and avoid the usual catastrophes within the core accretion model.

We note that the simulations in \cite{Rice:2004aa,Rice:2006aa} are not fully resolved vertically, indicating that pressure forces in the midplane are under-resolved (particularly at the thickness of the thinner dust layer). The resolution scale is of order the disc semi-thickness $H$, and is also of order the spiral arm width, and so on smaller scales pressure variations are smoothed out and weakened. This suggests that the effects reported there may be underestimates as the dust traps would be deeper at higher resolution. New \note{global} simulations along these lines with modern day resolution would be very useful. We note that dust self-gravity was included in \cite{Rice:2006aa}, which is necessary as the dust density can reach as high as the gas density, at which point it can also begin to collect gas. \note{In a series of papers \cite{Gibbons:2012aa,Gibbons:2014aa,Gibbons:2015aa} provide shearing-sheet simulations of self-gravitating gas with dust particles using the {\sc pencil} code, finding that self-gravitationally induced density waves are excellent sites for the rapid formation of large planetesimals.}

In summary, if dust is present early on (and we have no reason to believe it is not) then the self-gravitating gas in the form of spiral arms collects it into planetary embryos quickly. In this case, the 10 cm -- 1 m problems of the core accretion scenario are bypassed. Thus planetary cores, and gas giants, can form rapidly. At the same time, they may be repeatedly swept into the centre. In this picture the planet formation process is rapid, and inefficient. Most planets are lost to the central star and what we end up with as the observed set of planets around a star is in fact  the final stage of the process -- the last ones standing \citep[cf.][\note{see also \citealt{Baruteau:2011aa}}]{Armitage:2002aa}. The fast formation of a planet with mass $\gtrsim M_{\rm Jup}$ sufficient to open a gap also helps to alleviate the Type-I migration problem. The usual planet formation barriers can be overcome if planet formation begins early enough, during which the disc is massive enough and the planetesimal formation process is vigorous enough.

\section{Conclusions}
Current planet formation models are successful in explaining a wide variety of complex observed data, but remain incomplete. In this work we have argued  that the initial properties of protoplanetary discs need to be discussed with in the context of the star formation process. In all cases, whether the discs are formed either via monolithic collapse of a single core or from more realistic, dynamical star formation calculations, they are always initially massive enough to be self-gravitating with $M_{\rm disc} = (0.1-{\rm a~few})M_\star$. This implies that self-gravity may play an important role in the formation of planets, and that planet formation can be initiated at an early stage when the discs are young, and the stars are still contained in their nascent cocoons.

In these young, massive discs, strong, transient spiral arms are produced in the gas. These act as efficient dust traps, gathering together a mass of dust which is locally unstable to gravitational fragmentation (while the gas remains gravitationally stable to fragmentation). This dust fragmentation early in the disc lifetime can populate the disc with large planetesimals ($\gtrsim 1$\,m) and thus overcome the main barriers to the core accretion model. Planet formation can then continue unobstructed, with small planetesimals growing into planetary cores, some of which accrete gas envelopes. A clear implication of this picture is that planet formation occurs vigorously and early -- implying that by the time the discs are observable and no longer cocooned in their envelope, planets are already carving the discs into shape. 

\section*{Acknowledgments}
CJN is supported by the Science and Technology Facilities Council (grant number ST/M005917/1). The Theoretical Astrophysics Group at the University of Leicester is supported by an STFC Consolidated Grant.

\bibliographystyle{mnrasshort}
\bibliography{nixon}

\begin{thebibliography}{}
\makeatletter
\relax
\def\mn@urlcharsother{\let\do\@makeother \do\$\do\&\do\#\do\^\do\_\do\%\do\~}
\def\mn@doi{\begingroup\mn@urlcharsother \@ifnextchar [ {\mn@doi@}
  {\mn@doi@[]}}
\def\mn@doi@[#1]#2{\def\@tempa{#1}\ifx\@tempa\@empty \href
  {http://dx.doi.org/#2} {doi:#2}\else \href {http://dx.doi.org/#2} {#1}\fi
  \endgroup}
\def\mn@eprint#1#2{\mn@eprint@#1:#2::\@nil}
\def\mn@eprint@arXiv#1{\href {http://arxiv.org/abs/#1} {{\tt arXiv:#1}}}
\def\mn@eprint@dblp#1{\href {http://dblp.uni-trier.de/rec/bibtex/#1.xml}
  {dblp:#1}}
\def\mn@eprint@#1:#2:#3:#4\@nil{\def\@tempa {#1}\def\@tempb {#2}\def\@tempc
  {#3}\ifx \@tempc \@empty \let \@tempc \@tempb \let \@tempb \@tempa \fi \ifx
  \@tempb \@empty \def\@tempb {arXiv}\fi \@ifundefined
  {mn@eprint@\@tempb}{\@tempb:\@tempc}{\expandafter \expandafter \csname
  mn@eprint@\@tempb\endcsname \expandafter{\@tempc}}}

\bibitem[\protect\citeauthoryear{{ALMA Partnership} et~al.,}{{ALMA Partnership}
  et~al.}{2015}]{ALMA-Partnership:2015aa}
{ALMA Partnership} et~al., 2015, \mn@doi [\apjl] {10.1088/2041-8205/808/1/L3},
  \href {http://adsabs.harvard.edu/abs/2015ApJ...808L...3A} {808, L3}

\bibitem[\protect\citeauthoryear{{Adams}, {Ruden}  \& {Shu}}{{Adams}
  et~al.}{1989}]{Adams:1989aa}
{Adams} F.~C.,  {Ruden} S.~P.,   {Shu} F.~H.,  1989, \mn@doi [\apj]
  {10.1086/168187}, \href {http://adsabs.harvard.edu/abs/1989ApJ...347..959A}
  {347, 959}

\bibitem[\protect\citeauthoryear{{Alfv{\'e}n} \& {Arrhenius}}{{Alfv{\'e}n} \&
  {Arrhenius}}{1970}]{Alfven:1970aa}
{Alfv{\'e}n} H.,  {Arrhenius} G.,  1970, \mn@doi [\apss] {10.1007/BF00651333},
  \href {http://adsabs.harvard.edu/abs/1970Ap%26SS...8..338A} {8, 338}

\bibitem[\protect\citeauthoryear{{Armitage} et~al.,}{{Armitage}
  et~al.}{2002}]{Armitage:2002aa}
{Armitage} P.~J.,  et~al., 2002, \mn@doi [\mnras]
  {10.1046/j.1365-8711.2002.05531.x}, \href
  {http://adsabs.harvard.edu/abs/2002MNRAS.334..248A} {334, 248}

\bibitem[\protect\citeauthoryear{{Baruteau}, {Meru}  \&
  {Paardekooper}}{{Baruteau} et~al.}{2011}]{Baruteau:2011aa}
{Baruteau} C.,  {Meru} F.,   {Paardekooper} S.-J.,  2011, \mn@doi [\mnras]
  {10.1111/j.1365-2966.2011.19172.x}, \href
  {http://adsabs.harvard.edu/abs/2011MNRAS.416.1971B} {416, 1971}

\bibitem[\protect\citeauthoryear{{Bate}}{{Bate}}{2009}]{Bate:2009aa}
{Bate} M.~R.,  2009, \mn@doi [\mnras] {10.1111/j.1365-2966.2008.14106.x}, \href
  {http://adsabs.harvard.edu/abs/2009MNRAS.392..590B} {392, 590}

\bibitem[\protect\citeauthoryear{{Bate}}{{Bate}}{2012}]{Bate:2012aa}
{Bate} M.~R.,  2012, \mn@doi [\mnras] {10.1111/j.1365-2966.2011.19955.x}, \href
  {http://adsabs.harvard.edu/abs/2012MNRAS.419.3115B} {419, 3115}

\bibitem[\protect\citeauthoryear{{Bate}}{{Bate}}{2018}]{Bate:2018aa}
{Bate} M.~R.,  2018, \mnras~in~press

\bibitem[\protect\citeauthoryear{{Bate}, {Bonnell}  \& {Bromm}}{{Bate}
  et~al.}{2003}]{Bate:2003aa}
{Bate} M.~R.,  {Bonnell} I.~A.,   {Bromm} V.,  2003, \mn@doi [\mnras]
  {10.1046/j.1365-8711.2003.06210.x}, \href
  {http://adsabs.harvard.edu/abs/2003MNRAS.339..577B} {339, 577}

\bibitem[\protect\citeauthoryear{{Bitsch} et~al.,}{{Bitsch}
  et~al.}{2015}]{Bitsch:2015aa}
{Bitsch} B.,  et~al., 2015, \mn@doi [\aap] {10.1051/0004-6361/201424964}, \href
  {http://adsabs.harvard.edu/abs/2015A%26A...575A..28B} {575, A28}

\bibitem[\protect\citeauthoryear{{Blum} \& {Wurm}}{{Blum} \&
  {Wurm}}{2008}]{Blum:2008aa}
{Blum} J.,  {Wurm} G.,  2008, \mn@doi [\araa]
  {10.1146/annurev.astro.46.060407.145152}, \href
  {http://adsabs.harvard.edu/abs/2008ARA%26A..46...21B} {46, 21}

\bibitem[\protect\citeauthoryear{{Boss}}{{Boss}}{1997}]{Boss:1997aa}
{Boss} A.~P.,  1997, \mn@doi [Science] {10.1126/science.276.5320.1836}, \href
  {http://adsabs.harvard.edu/abs/1997Sci...276.1836B} {276, 1836}

\bibitem[\protect\citeauthoryear{{Chiang} \& {Laughlin}}{{Chiang} \&
  {Laughlin}}{2013}]{Chiang:2013aa}
{Chiang} E.,  {Laughlin} G.,  2013, \mn@doi [\mnras] {10.1093/mnras/stt424},
  \href {http://adsabs.harvard.edu/abs/2013MNRAS.431.3444C} {431, 3444}

\bibitem[\protect\citeauthoryear{{Clarke} \& {Pringle}}{{Clarke} \&
  {Pringle}}{1991}]{Clarke:1991aa}
{Clarke} C.~J.,  {Pringle} J.~E.,  1991, \mn@doi [\mnras]
  {10.1093/mnras/249.4.588}, \href
  {http://adsabs.harvard.edu/abs/1991MNRAS.249..588C} {249, 588}

\bibitem[\protect\citeauthoryear{{Coleman}, {Papaloizou}  \&
  {Nelson}}{{Coleman} et~al.}{2017}]{Coleman:2017aa}
{Coleman} G.~A.~L.,  {Papaloizou} J.~C.~B.,   {Nelson} R.~P.,  2017, \mn@doi
  [\mnras] {10.1093/mnras/stx1297}, \href
  {http://adsabs.harvard.edu/abs/2017MNRAS.470.3206C} {470, 3206}

\bibitem[\protect\citeauthoryear{{Dobbs}, {Bonnell}  \& {Pringle}}{{Dobbs}
  et~al.}{2006}]{Dobbs:2006aa}
{Dobbs} C.~L.,  {Bonnell} I.~A.,   {Pringle} J.~E.,  2006, \mn@doi [\mnras]
  {10.1111/j.1365-2966.2006.10794.x}, \href
  {http://adsabs.harvard.edu/abs/2006MNRAS.371.1663D} {371, 1663}

\bibitem[\protect\citeauthoryear{{Dobbs}, {Burkert}  \& {Pringle}}{{Dobbs}
  et~al.}{2011}]{Dobbs:2011aa}
{Dobbs} C.~L.,  {Burkert} A.,   {Pringle} J.~E.,  2011, \mn@doi [\mnras]
  {10.1111/j.1365-2966.2011.18371.x}, \href
  {http://adsabs.harvard.edu/abs/2011MNRAS.413.2935D} {413, 2935}

\bibitem[\protect\citeauthoryear{{Dobbs}, {Pringle}  \& {Burkert}}{{Dobbs}
  et~al.}{2012}]{Dobbs:2012aa}
{Dobbs} C.~L.,  {Pringle} J.~E.,   {Burkert} A.,  2012, \mn@doi [\mnras]
  {10.1111/j.1365-2966.2012.21558.x}, \href
  {http://adsabs.harvard.edu/abs/2012MNRAS.425.2157D} {425, 2157}

\bibitem[\protect\citeauthoryear{{Duquennoy} \& {Mayor}}{{Duquennoy} \&
  {Mayor}}{1991}]{Duquennoy:1991aa}
{Duquennoy} A.,  {Mayor} M.,  1991, \aap, \href
  {http://adsabs.harvard.edu/abs/1991A%26A...248..485D} {248, 485}

\bibitem[\protect\citeauthoryear{{Durisen} et~al.,}{{Durisen}
  et~al.}{2007}]{Durisen:2007aa}
{Durisen} R.~H.,  et~al., 2007, Protostars and Planets V, \href
  {http://adsabs.harvard.edu/abs/2007prpl.conf..607D} {pp 607--622}

\bibitem[\protect\citeauthoryear{{Edgeworth}}{{Edgeworth}}{1949}]{Edgeworth:1949aa}
{Edgeworth} K.~E.,  1949, \mn@doi [\mnras] {10.1093/mnras/109.5.600}, \href
  {http://adsabs.harvard.edu/abs/1949MNRAS.109..600E} {109, 600}

\bibitem[\protect\citeauthoryear{{Forgan} et~al.,}{{Forgan}
  et~al.}{2017}]{Forgan:2017aa}
{Forgan} D.~H.,  et~al., 2017, preprint, \href
  {http://adsabs.harvard.edu/abs/2017arXiv171101133F} {} (\mn@eprint {arXiv}
  {1711.01133})

\bibitem[\protect\citeauthoryear{{Fressin} et~al.,}{{Fressin}
  et~al.}{2013}]{Fressin:2013aa}
{Fressin} F.,  et~al., 2013, \mn@doi [\apj] {10.1088/0004-637X/766/2/81}, \href
  {http://adsabs.harvard.edu/abs/2013ApJ...766...81F} {766, 81}

\bibitem[\protect\citeauthoryear{{Gibbons}, {Rice}  \&
  {Mamatsashvili}}{{Gibbons} et~al.}{2012}]{Gibbons:2012aa}
{Gibbons} P.~G.,  {Rice} W.~K.~M.,   {Mamatsashvili} G.~R.,  2012, \mn@doi
  [\mnras] {10.1111/j.1365-2966.2012.21731.x}, \href
  {http://adsabs.harvard.edu/abs/2012MNRAS.426.1444G} {426, 1444}

\bibitem[\protect\citeauthoryear{{Gibbons}, {Mamatsashvili}  \&
  {Rice}}{{Gibbons} et~al.}{2014}]{Gibbons:2014aa}
{Gibbons} P.~G.,  {Mamatsashvili} G.~R.,   {Rice} W.~K.~M.,  2014, \mn@doi
  [\mnras] {10.1093/mnras/stu809}, \href
  {http://adsabs.harvard.edu/abs/2014MNRAS.442..361G} {442, 361}

\bibitem[\protect\citeauthoryear{{Gibbons}, {Mamatsashvili}  \&
  {Rice}}{{Gibbons} et~al.}{2015}]{Gibbons:2015aa}
{Gibbons} P.~G.,  {Mamatsashvili} G.~R.,   {Rice} W.~K.~M.,  2015, \mn@doi
  [\mnras] {10.1093/mnras/stv1766}, \href
  {http://adsabs.harvard.edu/abs/2015MNRAS.453.4232G} {453, 4232}

\bibitem[\protect\citeauthoryear{{Hall}, {Forgan}  \& {Rice}}{{Hall}
  et~al.}{2017}]{Hall:2017aa}
{Hall} C.,  {Forgan} D.,   {Rice} K.,  2017, \mn@doi [\mnras]
  {10.1093/mnras/stx1244}, \href
  {http://adsabs.harvard.edu/abs/2017MNRAS.470.2517H} {470, 2517}

\bibitem[\protect\citeauthoryear{{Hall} et~al.,}{{Hall}
  et~al.}{2018}]{Hall:2018aa}
{Hall} C.,  et~al., 2018, \mn@doi [\mnras] {10.1093/mnras/sty550}, \href
  {http://adsabs.harvard.edu/abs/2018MNRAS.tmp..517H} {}

\bibitem[\protect\citeauthoryear{{Helled} \& {Bodenheimer}}{{Helled} \&
  {Bodenheimer}}{2010}]{Helled:2010aa}
{Helled} R.,  {Bodenheimer} P.,  2010, \mn@doi [\icarus]
  {10.1016/j.icarus.2009.11.023}, \href
  {http://adsabs.harvard.edu/abs/2010Icar..207..503H} {207, 503}

\bibitem[\protect\citeauthoryear{{Hopkins}}{{Hopkins}}{2016}]{Hopkins:2016aa}
{Hopkins} P.~F.,  2016, \mn@doi [\mnras] {10.1093/mnras/stv2820}, \href
  {http://adsabs.harvard.edu/abs/2016MNRAS.456.2383H} {456, 2383}

\bibitem[\protect\citeauthoryear{{Ilee} et~al.,}{{Ilee}
  et~al.}{2017}]{Ilee:2017aa}
{Ilee} J.~D.,  et~al., 2017, \mn@doi [\mnras] {10.1093/mnras/stx1966}, \href
  {http://adsabs.harvard.edu/abs/2017MNRAS.472..189I} {472, 189}

\bibitem[\protect\citeauthoryear{{Jin} \& {Sui}}{{Jin} \&
  {Sui}}{2010}]{Jin:2010aa}
{Jin} L.,  {Sui} N.,  2010, \mn@doi [\apj] {10.1088/0004-637X/710/2/1179},
  \href {http://adsabs.harvard.edu/abs/2010ApJ...710.1179J} {710, 1179}

\bibitem[\protect\citeauthoryear{{Kenyon} et~al.,}{{Kenyon}
  et~al.}{1990}]{Kenyon:1990aa}
{Kenyon} S.~J.,  et~al., 1990, \mn@doi [\aj] {10.1086/115380}, \href
  {http://adsabs.harvard.edu/abs/1990AJ.....99..869K} {99, 869}

\bibitem[\protect\citeauthoryear{{Kratter} \& {Lodato}}{{Kratter} \&
  {Lodato}}{2016}]{Kratter:2016aa}
{Kratter} K.,  {Lodato} G.,  2016, \mn@doi [\araa]
  {10.1146/annurev-astro-081915-023307}, \href
  {http://adsabs.harvard.edu/abs/2016ARA%26A..54..271K} {54, 271}

\bibitem[\protect\citeauthoryear{{Kratter} \& {Matzner}}{{Kratter} \&
  {Matzner}}{2006}]{Kratter:2006aa}
{Kratter} K.~M.,  {Matzner} C.~D.,  2006, \mn@doi [\mnras]
  {10.1111/j.1365-2966.2006.11103.x}, \href
  {http://adsabs.harvard.edu/abs/2006MNRAS.373.1563K} {373, 1563}

\bibitem[\protect\citeauthoryear{{Krumholz}, {Klein}  \& {McKee}}{{Krumholz}
  et~al.}{2011}]{Krumholz:2011aa}
{Krumholz} M.~R.,  {Klein} R.~I.,   {McKee} C.~F.,  2011, \mn@doi [\apj]
  {10.1088/0004-637X/740/2/74}, \href
  {http://adsabs.harvard.edu/abs/2011ApJ...740...74K} {740, 74}

\bibitem[\protect\citeauthoryear{{K{\"u}ffmeier}, {Haugb{\o}lle}  \&
  {Nordlund}}{{K{\"u}ffmeier} et~al.}{2017}]{Kuffmeier:2017aa}
{K{\"u}ffmeier} M.,  {Haugb{\o}lle} T.,   {Nordlund} {\AA}.,  2017, \mn@doi
  [\apj] {10.3847/1538-4357/aa7c64}, \href
  {http://adsabs.harvard.edu/abs/2017ApJ...846....7K} {846, 7}

\bibitem[\protect\citeauthoryear{{Kuiper}}{{Kuiper}}{1956}]{Kuiper:1956aa}
{Kuiper} G.~P.,  1956, \jrasc, \href
  {http://adsabs.harvard.edu/abs/1956JRASC..50..158K} {50, 158}

\bibitem[\protect\citeauthoryear{{Kusaka}, {Nakano}  \& {Hayashi}}{{Kusaka}
  et~al.}{1970}]{Kusaka:1970aa}
{Kusaka} T.,  {Nakano} T.,   {Hayashi} C.,  1970, \mn@doi [Progress of
  Theoretical Physics] {10.1143/PTP.44.1580}, \href
  {http://adsabs.harvard.edu/abs/1970PThPh..44.1580K} {44, 1580}

\bibitem[\protect\citeauthoryear{{Laskar}}{{Laskar}}{1996}]{Laskar:1996aa}
{Laskar} J.,  1996, \mn@doi [Celestial Mechanics and Dynamical Astronomy]
  {10.1007/BF00051610}, \href
  {http://adsabs.harvard.edu/abs/1996CeMDA..64..115L} {64, 115}

\bibitem[\protect\citeauthoryear{{Lecar} \& {Franklin}}{{Lecar} \&
  {Franklin}}{1973}]{Lecar:1973aa}
{Lecar} M.,  {Franklin} F.~A.,  1973, \mn@doi [\icarus]
  {10.1016/0019-1035(73)90015-8}, \href
  {http://adsabs.harvard.edu/abs/1973Icar...20..422L} {20, 422}

\bibitem[\protect\citeauthoryear{{Lin} \& {Pringle}}{{Lin} \&
  {Pringle}}{1987}]{Lin:1987aa}
{Lin} D.~N.~C.,  {Pringle} J.~E.,  1987, \mn@doi [\mnras]
  {10.1093/mnras/225.3.607}, \href
  {http://adsabs.harvard.edu/abs/1987MNRAS.225..607L} {225, 607}

\bibitem[\protect\citeauthoryear{{Lin} \& {Pringle}}{{Lin} \&
  {Pringle}}{1990}]{Lin:1990aa}
{Lin} D.~N.~C.,  {Pringle} J.~E.,  1990, \mn@doi [\apj] {10.1086/169004}, \href
  {http://adsabs.harvard.edu/abs/1990ApJ...358..515L} {358, 515}

\bibitem[\protect\citeauthoryear{{Lines} et~al.,}{{Lines}
  et~al.}{2015}]{Lines:2015aa}
{Lines} S.,  et~al., 2015, \mn@doi [\aap] {10.1051/0004-6361/201526295}, \href
  {http://adsabs.harvard.edu/abs/2015A%26A...582A...5L} {582, A5}

\bibitem[\protect\citeauthoryear{{Lissauer}}{{Lissauer}}{1993}]{Lissauer:1993aa}
{Lissauer} J.~J.,  1993, \mn@doi [\araa] {10.1146/annurev.aa.31.090193.001021},
  \href {http://adsabs.harvard.edu/abs/1993ARA%26A..31..129L} {31, 129}

\bibitem[\protect\citeauthoryear{{Martin} \& {Lubow}}{{Martin} \&
  {Lubow}}{2011}]{Martin:2011aa}
{Martin} R.~G.,  {Lubow} S.~H.,  2011, \mn@doi [\apjl]
  {10.1088/2041-8205/740/1/L6}, \href
  {http://adsabs.harvard.edu/abs/2011ApJ...740L...6M} {740, L6}

\bibitem[\protect\citeauthoryear{{Martin} et~al.,}{{Martin}
  et~al.}{2012}]{Martin:2012aa}
{Martin} R.~G.,  et~al., 2012, \mn@doi [\mnras]
  {10.1111/j.1365-2966.2012.21076.x}, \href
  {http://adsabs.harvard.edu/abs/2012MNRAS.423.2718M} {423, 2718}

\bibitem[\protect\citeauthoryear{{Meru} et~al.,}{{Meru}
  et~al.}{2017}]{Meru:2017aa}
{Meru} F.,  et~al., 2017, \mn@doi [\apjl] {10.3847/2041-8213/aa6837}, \href
  {http://adsabs.harvard.edu/abs/2017ApJ...839L..24M} {839, L24}

\bibitem[\protect\citeauthoryear{{Mizuno}}{{Mizuno}}{1980}]{Mizuno:1980aa}
{Mizuno} H.,  1980, \mn@doi [Progress of Theoretical Physics]
  {10.1143/PTP.64.544}, \href
  {http://adsabs.harvard.edu/abs/1980PThPh..64..544M} {64, 544}

\bibitem[\protect\citeauthoryear{{Mutter}, {Pierens}  \& {Nelson}}{{Mutter}
  et~al.}{2017}]{Mutter:2017aa}
{Mutter} M.~M.,  {Pierens} A.,   {Nelson} R.~P.,  2017, \mn@doi [\mnras]
  {10.1093/mnras/stw2768}, \href
  {http://adsabs.harvard.edu/abs/2017MNRAS.465.4735M} {465, 4735}

\bibitem[\protect\citeauthoryear{{Nakamoto} \& {Nakagawa}}{{Nakamoto} \&
  {Nakagawa}}{1995}]{Nakamoto:1995aa}
{Nakamoto} T.,  {Nakagawa} Y.,  1995, \mn@doi [\apj] {10.1086/175697}, \href
  {http://adsabs.harvard.edu/abs/1995ApJ...445..330N} {445, 330}

\bibitem[\protect\citeauthoryear{{Nayakshin}}{{Nayakshin}}{2017}]{Nayakshin:2017aa}
{Nayakshin} S.,  2017, \mn@doi [\pasa] {10.1017/pasa.2016.55}, \href
  {http://adsabs.harvard.edu/abs/2017PASA...34....2N} {34, e002}

\bibitem[\protect\citeauthoryear{{P{\'e}rez} et~al.,}{{P{\'e}rez}
  et~al.}{2016}]{Perez:2016aa}
{P{\'e}rez} L.~M.,  et~al., 2016, \mn@doi [Science] {10.1126/science.aaf8296},
  \href {http://adsabs.harvard.edu/abs/2016Sci...353.1519P} {353, 1519}

\bibitem[\protect\citeauthoryear{{Pollack} et~al.,}{{Pollack}
  et~al.}{1996}]{Pollack:1996aa}
{Pollack} J.~B.,  et~al., 1996, \mn@doi [\icarus] {10.1006/icar.1996.0190},
  \href {http://adsabs.harvard.edu/abs/1996Icar..124...62P} {124, 62}

\bibitem[\protect\citeauthoryear{{Pringle}}{{Pringle}}{1981}]{Pringle:1981aa}
{Pringle} J.~E.,  1981, \mn@doi [\araa] {10.1146/annurev.aa.19.090181.001033},
  \href {http://adsabs.harvard.edu/abs/1981ARA%26A..19..137P} {19, 137}

\bibitem[\protect\citeauthoryear{{Pringle}}{{Pringle}}{1989}]{Pringle:1989aa}
{Pringle} J.~E.,  1989, \mn@doi [\mnras] {10.1093/mnras/239.2.361}, \href
  {http://adsabs.harvard.edu/abs/1989MNRAS.239..361P} {239, 361}

\bibitem[\protect\citeauthoryear{{Pringle}}{{Pringle}}{1991}]{Pringle:1991aa}
{Pringle} J.~E.,  1991, in {Lada} C.~J.,  {Kylafis} N.~D.,  eds,  NATO Advanced
  Science Institutes (ASI) Series C Vol. 342, NATO Advanced Science Institutes
  (ASI) Series C. p.~437

\bibitem[\protect\citeauthoryear{{Rafikov}}{{Rafikov}}{2005}]{Rafikov:2005aa}
{Rafikov} R.~R.,  2005, \mn@doi [\apjl] {10.1086/428899}, \href
  {http://adsabs.harvard.edu/abs/2005ApJ...621L..69R} {621, L69}

\bibitem[\protect\citeauthoryear{{Raghavan} et~al.,}{{Raghavan}
  et~al.}{2010}]{Raghavan:2010aa}
{Raghavan} D.,  et~al., 2010, \mn@doi [\apjs] {10.1088/0067-0049/190/1/1},
  \href {http://adsabs.harvard.edu/abs/2010ApJS..190....1R} {190, 1}

\bibitem[\protect\citeauthoryear{{Rice} et~al.,}{{Rice}
  et~al.}{2004}]{Rice:2004aa}
{Rice} W.~K.~M.,  et~al., 2004, \mn@doi [\mnras]
  {10.1111/j.1365-2966.2004.08339.x}, \href
  {http://adsabs.harvard.edu/abs/2004MNRAS.355..543R} {355, 543}

\bibitem[\protect\citeauthoryear{{Rice} et~al.,}{{Rice}
  et~al.}{2006}]{Rice:2006aa}
{Rice} W.~K.~M.,  et~al., 2006, \mn@doi [\mnras]
  {10.1111/j.1745-3933.2006.00215.x}, \href
  {http://adsabs.harvard.edu/abs/2006MNRAS.372L...9R} {372, L9}

\bibitem[\protect\citeauthoryear{{Rice}, {Mayo}  \& {Armitage}}{{Rice}
  et~al.}{2010}]{Rice:2010aa}
{Rice} W.~K.~M.,  {Mayo} J.~H.,   {Armitage} P.~J.,  2010, \mn@doi [\mnras]
  {10.1111/j.1365-2966.2009.15992.x}, \href
  {http://adsabs.harvard.edu/abs/2010MNRAS.402.1740R} {402, 1740}

\bibitem[\protect\citeauthoryear{{Safronov}}{{Safronov}}{1967}]{Safronov:1967aa}
{Safronov} V.~S.,  1967, \sovast, \href
  {http://adsabs.harvard.edu/abs/1967SvA....10..650S} {10, 650}

\bibitem[\protect\citeauthoryear{{Shakura} \& {Sunyaev}}{{Shakura} \&
  {Sunyaev}}{1973}]{Shakura:1973aa}
{Shakura} N.~I.,  {Sunyaev} R.~A.,  1973, \aap, \href
  {http://adsabs.harvard.edu/abs/1973A%26A....24..337S} {24, 337}

\bibitem[\protect\citeauthoryear{{Shu}, {Adams}  \& {Lizano}}{{Shu}
  et~al.}{1987}]{Shu:1987aa}
{Shu} F.~H.,  {Adams} F.~C.,   {Lizano} S.,  1987, \mn@doi [\araa]
  {10.1146/annurev.aa.25.090187.000323}, \href
  {http://adsabs.harvard.edu/abs/1987ARA%26A..25...23S} {25, 23}

\bibitem[\protect\citeauthoryear{{Simon} et~al.,}{{Simon}
  et~al.}{2015}]{Simon:2015aa}
{Simon} J.~B.,  et~al., 2015, \mn@doi [\mnras] {10.1093/mnras/stv2070}, \href
  {http://adsabs.harvard.edu/abs/2015MNRAS.454.1117S} {454, 1117}

\bibitem[\protect\citeauthoryear{{Stamatellos}}{{Stamatellos}}{2015}]{Stamatellos:2015aa}
{Stamatellos} D.,  2015, \mn@doi [\apjl] {10.1088/2041-8205/810/1/L11}, \href
  {http://adsabs.harvard.edu/abs/2015ApJ...810L..11S} {810, L11}

\bibitem[\protect\citeauthoryear{{Thorngren} et~al.,}{{Thorngren}
  et~al.}{2016}]{Thorngren:2016aa}
{Thorngren} D.~P.,  et~al., 2016, \mn@doi [\apj] {10.3847/0004-637X/831/1/64},
  \href {http://adsabs.harvard.edu/abs/2016ApJ...831...64T} {831, 64}

\bibitem[\protect\citeauthoryear{{Tobin} et~al.,}{{Tobin}
  et~al.}{2012}]{Tobin:2012aa}
{Tobin} J.~J.,  et~al., 2012, \mn@doi [\nat] {10.1038/nature11610}, \href
  {http://adsabs.harvard.edu/abs/2012Natur.492...83T} {492, 83}

\bibitem[\protect\citeauthoryear{{Tobin} et~al.,}{{Tobin}
  et~al.}{2015}]{Tobin:2015aa}
{Tobin} J.~J.,  et~al., 2015, \mn@doi [\apj] {10.1088/0004-637X/805/2/125},
  \href {http://adsabs.harvard.edu/abs/2015ApJ...805..125T} {805, 125}

\bibitem[\protect\citeauthoryear{{Tobin} et~al.,}{{Tobin}
  et~al.}{2016}]{Tobin:2016aa}
{Tobin} J.~J.,  et~al., 2016, \mn@doi [\nat] {10.1038/nature20094}, \href
  {http://adsabs.harvard.edu/abs/2016Natur.538..483T} {538, 483}

\bibitem[\protect\citeauthoryear{{Tomida} et~al.,}{{Tomida}
  et~al.}{2017}]{Tomida:2017aa}
{Tomida} K.,  et~al., 2017, \mn@doi [\apjl] {10.3847/2041-8213/835/1/L11},
  \href {http://adsabs.harvard.edu/abs/2017ApJ...835L..11T} {835, L11}

\bibitem[\protect\citeauthoryear{{Walch} et~al.,}{{Walch}
  et~al.}{2009}]{Walch:2009aa}
{Walch} S.,  et~al., 2009, \mn@doi [\mnras] {10.1111/j.1365-2966.2009.15293.x},
  \href {http://adsabs.harvard.edu/abs/2009MNRAS.400...13W} {400, 13}

\bibitem[\protect\citeauthoryear{{Weidenschilling}}{{Weidenschilling}}{1977a}]{Weidenschilling:1977ab}
{Weidenschilling} S.~J.,  1977a, \mn@doi [\apss] {10.1007/BF00642464}, \href
  {http://adsabs.harvard.edu/abs/1977Ap%26SS..51..153W} {51, 153}

\bibitem[\protect\citeauthoryear{{Weidenschilling}}{{Weidenschilling}}{1977b}]{Weidenschilling:1977aa}
{Weidenschilling} S.~J.,  1977b, \mn@doi [\mnras] {10.1093/mnras/180.1.57},
  \href {http://adsabs.harvard.edu/abs/1977MNRAS.180...57W} {180, 57}

\bibitem[\protect\citeauthoryear{{Weidenschilling} \&
  {Cuzzi}}{{Weidenschilling} \& {Cuzzi}}{1993}]{Weidenschilling:1993aa}
{Weidenschilling} S.~J.,  {Cuzzi} J.~N.,  1993, in {Levy} E.~H.,  {Lunine}
  J.~I.,  eds, Protostars and Planets III. pp 1031--1060

\bibitem[\protect\citeauthoryear{{Winn} \& {Fabrycky}}{{Winn} \&
  {Fabrycky}}{2015}]{Winn:2015aa}
{Winn} J.~N.,  {Fabrycky} D.~C.,  2015, \mn@doi [\araa]
  {10.1146/annurev-astro-082214-122246}, \href
  {http://adsabs.harvard.edu/abs/2015ARA%26A..53..409W} {53, 409}

\bibitem[\protect\citeauthoryear{{Yen} et~al.,}{{Yen}
  et~al.}{2015}]{Yen:2015aa}
{Yen} H.-W.,  et~al., 2015, \mn@doi [\apj] {10.1088/0004-637X/799/2/193}, \href
  {http://adsabs.harvard.edu/abs/2015ApJ...799..193Y} {799, 193}

\bibitem[\protect\citeauthoryear{{Zinzi} \& {Turrini}}{{Zinzi} \&
  {Turrini}}{2017}]{Zinzi:2017aa}
{Zinzi} A.,  {Turrini} D.,  2017, \mn@doi [\aap] {10.1051/0004-6361/201731595},
  \href {http://adsabs.harvard.edu/abs/2017A%26A...605L...4Z} {605, L4}

\makeatother
\end{thebibliography}

\bsp
\label{lastpage}
\end{document}